\documentclass{article}
\usepackage[final]{graphics}
\usepackage{psfig}

\newcommand{\be}{\begin{equation}}
\newcommand{\ee}{\end{equation}}
\newcommand{\ba}{\begin{eqnarray}}
\newcommand{\ea}{\end{eqnarray}}

\begin{document}
\renewcommand{\figurename}{{\bf Fig.}}
\renewcommand{\tablename}{{\bf Tab.}}

\title{Finite temperature quantum correlations in $SU(2)_c$ quark states
  and quantum spin models}

\author{S.~Hamieh$^1$ and A.~Tawfik$^2$\\[0.3cm]
     {\small\it $^1$Kernfysisch Versneller Instituut, Zernikelaan 25,
                9747~AA~Groningen, The Netherlands}\\ 
     {\small\it  $^2$Fakult{\"a}t f{\"u}r Physik, Universit{\"a}t Bielefeld,
                Postfach~100131, D-33501 Bielefeld, Germany}
}

\date{}
\maketitle

\begin{abstract}
The entanglement at finite temperatures are analyzed by using thermal
models for colored quarks making up the hadron physical states. We have
found that these quantum correlations entirely vanishes at \hbox{$T_c\geq
  m_q/\ln(1.5)$}. For temperatures larger than $T_c$ the correlations are
classical. Also we worked out the entanglement for the transverse
Ising spin chain. In dependence on both temperature $T$ and transverse
field $\lambda$ we can identify a certain region, where the quantum effects
are likely to dominate the system. We suggest the mutual 
information as a quantitative measure for the correlations in ground
state.

\end{abstract}

\section{Introduction} 
\label{sect:1}

One of the consequences of existing finite entropy at
zero temperature is the understanding of Gibbs paradox, which represents one of
very old thermodynamical problems. We now know that one has to take
into consider the distinguishability for better understanding the finite
entropy in ground state. On one hand side, we find that the change in the  
entropy, when non-identical particles have been mixed at fixed
particle number and volume, goes to zero for vanishing
temperature. This is known as the Nernst's heat theorem. On the other hand,
for identical particles the entropy goes to a 
temperature-independent value ${\cal O}(n)$ proportional to the
number of mixed states at zero temperature. It is obviously
independent on all thermodynamical quantities other than the number of
particles itself. This is usually stated under the name of the 
third low of thermodynamics and dates back to the beginning of last
century. For a non-comprehensible reason one used to mix up these two cases
and suppose that in all systems the entropy goes to zero for $T\rightarrow
0$. This might be well founded from the accommodativeness to 
approximate $S$ to zero for zero temperature.  

There are many physical systems, in which one has to avoid this
approximation. For instance, Schr{\"o}dinger expected that the ground state
of a gas of atoms encloses $2^N$ degenerate configurations in its 
structure, which must then provide for it an entropy of $N\ln2$. The first
application of these ideas on the quark matter has been introduced
in~\cite{Miller}. It has been found that the $SU(2)_c$ color symmetry for
each of the colored quarks in the ground state gives an entropy
$\ln2$. This has been extended to models at finite temperatures
in~\cite{Mill03}. Such an entropy can be seen as a reflection of the
probabilities of quark mixing maneuver inside the mesons. Therefore, at
zero temperature the confined quarks seem to be continuously tousled
objects.  
    
A further example for the physical systems, in which a finite entropy
 exists at zero temperature is the recent 
lattice QCD results for the equation of state of
quark-antiquark~\cite{KMTZ}. It has been found that the entropy
takes the constant value $2\ln3$ for $T\rightarrow 0$. This is obviously
 well compatible with the evaluation of the entropy of $SU(3)_c$ colored 
singlet state based on 
the models given in~\cite{Mill03}. There are other applications, for
instance, the equation of state of the hadronic matter at very low 
temperatures~\cite{Mill032}, the quark-pair condensates at very high
densities and low temperatures and the compact stellar
objects~\cite{Mill033}.

In this paper we study the quantum correlations and their impacts in
systems with $SU(2)_c$ color symmetry at finite 
temperatures. We suggest the concurrence as a measure for the thermal
effects on these correlations. We aim for evaluating a temperature, up
which the quantum correlations can survive. In section~\ref{sec:3} we make
a look at the transverse Ising model. The conclusions and outlook are given in
section~\ref{sec:4}

\section{$SU(2)_c$ quarks states} 
\label{sec:2}

\begin{figure}[htb]
\centerline{\psfig{width=10cm,figure=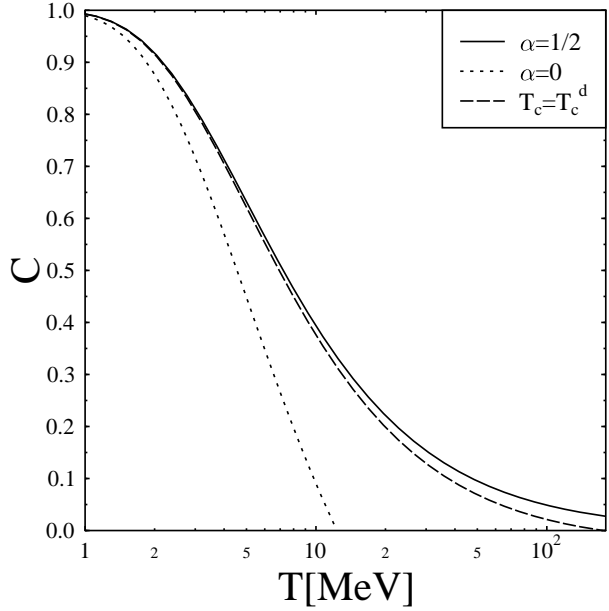}}
\vspace{-1.5cm}\caption{ Concurrence ${\cal C}$ as a function of $T$ for
  two-site correlations in $SU(2)$ quark states  with
  $m_q=5$ MeV. For $\alpha = 0$, the `critical' temperature takes the value
  $T\sim 12.5$ MeV (dotted line). For $\alpha \rightarrow 1/2$ we find that
  $T_c\rightarrow\infty$ (solid line), meanwhile for $\alpha =0.499$ we
  get $T_c=T_c^d$ (dashed line).
\protect\label{phatra}}
\end{figure}

Using the models proposed in~\cite{Mill03} based on the Boltzmann weighting
for the finite temperature states of colored quarks we postulate that the
total density matrix in the ground state of a meson 
singlet-state of $SU(2)_c$ color symmetry in the Hilbert-Schmidt space
$C^2\otimes C^2$ could be written as 
\be 
\rho_{AB}={1\over 4}\left(I\otimes I -\alpha
  e^{-\frac{\epsilon(p)}{T}}({\sigma_z}\otimes I +I \otimes { \sigma_z})
  -\left(1- e^{-\frac{\epsilon(p)}{T}}\right)
  \sum_{i=1}^3\sigma_i\otimes\sigma_i\right  
  ) \label{eq3} 
\ee
$I$ stands for the identity operator, $\epsilon(p)=\sqrt{m^2+p^2}$ is
the single particle relative energy, $\sigma^i; i=x,y,z$ are the standard
Pauli matrices and the quantity $\alpha e^{-\epsilon(p)/T}$ describes the
evolution of the color as a function of the temperature $T$. $\alpha$
represents the free parameter in this model.  
Clearly, for $\alpha=1$ we go back to the expression for the thermal
reduced density matrix given in~\cite{Mill03} in equation~(10). However, in
order to fulfill the condition that the total density matrix has a
non-negative value, we firstly suppose that  
\be 
-\frac{1}{2}\leq\alpha\leq \frac{1}{2}
\ee
For $\alpha=0$, the mixed state in the whole system will be
colorless for all temperatures. At $T=0$ equation~\ref{eq3} gives the
expected density matrix for the pure colored singlet state in $SU(2)_c$ ground
state. As introduced above, Eq.~\ref{eq3} is also able to reproduce the
finite temperature behavior, where we expect
no quantum correlations to be exist. There is no direct description of a
kind of quantum transition available by means of Eq.~\ref{eq3}. One
should notice that $\rho_{AB}$ satisfy the
condition ${\rm Tr}\;\rho^2\leq 1$.\\

\begin{figure}[htb]
\centerline{\psfig{width=10cm,figure=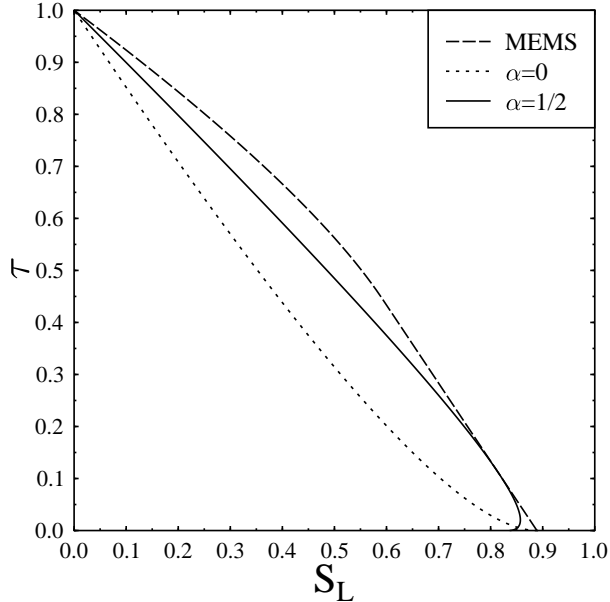}}
\vspace{-2cm}\caption{$\tau\equiv{\cal C}^2$ as a function of the linear
  entropy 
  $S_L=4/3 (1-{\rm Tr }\rho^2)$ for the Maximally~Entangled~Mixed~States
  (MEMS) and for the entropy calculated for $\alpha=0$ and $1/2$. 
\protect\label{entent}}
\end{figure}

 In order to compute the entanglement of such states at finite
 temperatures, we use the concurrence
 formalism~\cite{Benn96}, which is  defined as 
\be
{\cal C}=\max\{\lambda_1-\lambda_2-\lambda_3-\lambda_4,0\} \label{eq:concur1}
\ee 
where $\lambda$'s are the square roots of the eigenvalues in
 decreasing order of $\rho_{AB}(\sigma_y\otimes\sigma_y\rho_{AB}^{\star}
 \sigma_y\otimes\sigma_y)$. $\rho_{AB}^{\star}$ is the corresponding
 complex conjugation in the computational basis  $\{|++\rangle,
 |+-\rangle, |-+\rangle,|--\rangle\}$. After straightforward little algebra we
 find that  
\be 
{{\cal
 C}(T)}=\left\{\begin{array}{c}1 - (1+\frac{\sqrt{1 -
 4\alpha^2}}{2})e^{-\frac{\epsilon(p)}{T}}\quad\;\; {\rm if} \;\;T\leq T_c\\ \\
0\quad \quad\quad\quad \quad\quad\quad\quad \quad\quad \quad{\rm
 if}\,\,T\geq T_c 
\end{array}\right.
\ee
where $\frac{\epsilon(p)}{\ln(3/2)}\leq T_c$. \\

The value of $\alpha$ can then be fixed through the hypothetical assumption
that the transition temperature $T_c^d$ from {\it confined} hadron 
gas to {\it deconfined} quark gluon phase at zero chemical potential
would have the same value as that for the transition 
from {\it entangled} to {\it non-entangled} $SU(2)_c$ colored states. In
this case 
\be
\alpha^2=\frac{1}{4}\left(1-4\left(e^{\frac{m_q}{T_c^d}}-1\right)^2\right)
\ee
From the lattice QCD simulations we know that for two quark
flavors the transition temperature at $\mu_q=0$ is $\sim 173$ MeV. If we
take the mass threshold $m_q(\mu)$ of light quarks as
$\epsilon(p)$~\cite{Sala00} where $\mu$ in this case is a renormalization
group scale and for simplicity we 
set $m_q(\mu(T))=5\,MeV$, we therefore find that $\alpha$ for
which $T_c=T_c^d$ is $0.499$. 

As shortly given above, the entanglement  transition  is
corresponding to 
${\cal C}=0$. Therefrom, we can simply estimate $T_c$  for the
simple case that $\alpha=0$ and $m_q=5\;$MeV, as
\be 
T_c\sim 12.5\, MeV\ll T_c^d
\ee
This represents the lower value of $T_c$.    
In Fig.~\ref{phatra} we draw ${\cal C}$ for the two-site
correlations in $SU(2)_c$ quark states as a function of
$T$ with different values of $\alpha$. 
We find that $T_c \ll T_c^d$ for $\alpha=0$, meanwhile $T_c \rightarrow
\infty$ for $\alpha\rightarrow 0.5$. \\   

\begin{figure}[htb]
\centerline{\psfig{width=10cm,figure=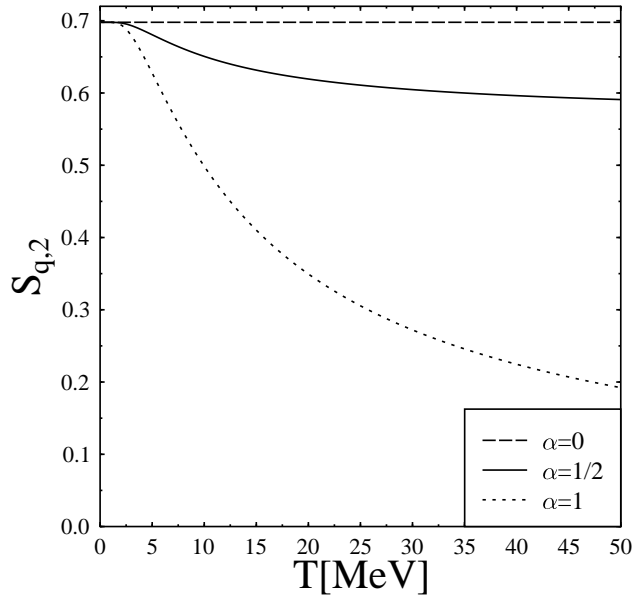}}
\vspace{-2cm}
\caption{Single-site color entropy as a function of the temperature $T$ (see
  text for more details). 
\protect\label{ent}}
\end{figure}

The density matrices given in Eq.~\ref{eq3} have some particular
properties. For specific values of $T$ and $\alpha$ (see Fig.~\ref{entent})
they belong to the states with high entanglement, the so-called
Maximally~Entangled~Mixed~States (MEMS)~\cite{Munr01}. This would suggest
that the thermal evolution of the colored quark will maximize the
entanglement at given degree of mixing. 

Moreover, it has been shown in~\cite{Woot01} that in
an antiferromagnetic ring for Heisenberg model (one-dimensional with
boundary conditions) with an even number of spin-$1/2$ the maximum
entanglement is equal to the entanglement of the ground state (minimum
energy) with zero $z$ component of the spin. The transverse spin models
will be studied in section~\ref{sec:3}. Therefore, we do not need for the
moment to elaborate any further details. 

In Fig.~\ref{entent} we plot $\tau\equiv C^2$
versus the linear entropy \hbox{$S_L=\frac{4}{3}(1-{\rm Tr}\rho^2)$}. Clearly,
we notice that the states with the properties given in Eq.~\ref{eq3}
typically do not maximize the entanglement at a given degree of state
mixing. These are the states with $\alpha=0$. But for the special case
$\alpha\rightarrow 0.5$ such 
states are very close to maximize the entanglement at all degrees of mixing.  
For $S_L=0$, i.e., pure state, the results for MEMS and all values of
$\alpha$ are coincident. \\

\begin{figure}[htb]
\centerline{\psfig{width=10cm,figure=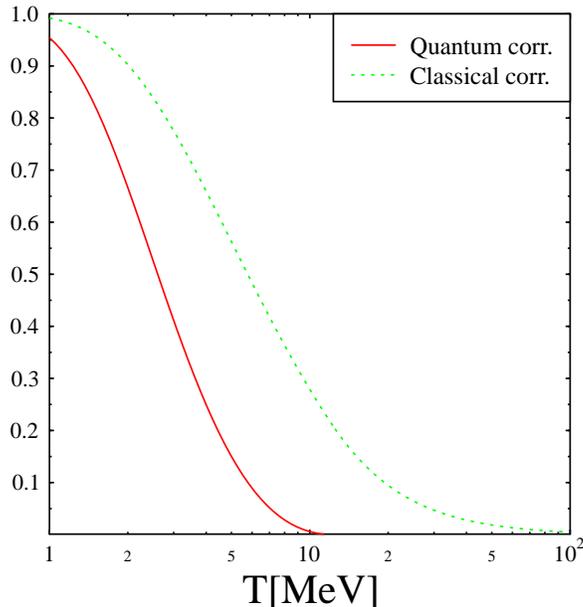}}
\vspace{-1cm}\caption{The relative entropy of entanglement is depicted in
  dependence on $T$ ($\alpha=0$ and $m_q=5\;$MeV). The quantum correlations are
  given in Eq.~\ref{eq:ccorr}.
 The classical
  correlations are the difference between the total correlations from  
the von~Neumann mutual information and the quantum correlations. 
At $T_c$ the quantum correlations are entirely vanishing meanwhile the
classical ones remain finite for all temperatures. 
\protect\label{entcla}}
\end{figure}

As shortly given in section~\ref{sect:1}, the effects of the quantum entropy
 of $SU(2)_c$ of colored quark states on 
 the bag pressure have been discussed in~\cite{Mill032}. This we can also be
 analyzed here by computing the color entropy of single-quark in one meson. In
 Fig.~\ref{ent} we show the results for different values of $\alpha$. In
 the case $\alpha=0$ we notice that the quark quantum entropy $S_{q,2}$ is a
 constant for all temperatures. As given in~\cite{Mill032} this case is
 corresponding to excluding the quantum entropy from the equation of
 state. The reason for this result is clear, since for
$\alpha=0$ all states are composed of completely random {\it sub}states
mixed together with singlet states. In both groups the single-site
entropy is maximum (the net color is zero). Though
 when $\alpha=0$ there is no color polarization for each 
 quark (local polarization is zero) and the overall color is also zero, the
 colors of single-quark are still correlated. For  
temperatures below $T_c$ the correlations is a mixture of quantum and
classical ones. However, for $T>T_c$ the correlations are completely
classical. To illustrate these features we plot in Fig.~\ref{entcla} the
 entanglement measured by the relative entropy from the entanglement, which is
 defined as~\cite{Vedr97} 
\be 
E_r=\min_{\rho^{*}\in {\cal D}}S(\rho\|\rho^{*}), \label{eq:ccorr}
\ee
For completeness the discussion, we mention that the classical correlations
are defined  as the difference between the total correlations measured by
the von~Neumann mutual information and the quantum correlations measured by
the relative entropy from the entanglement~\cite{Hami03}.  
\be 
\Psi(\rho)=S(\rho\|\rho_A\otimes\rho_B)-  \min_{\rho^{*}\in {\cal
    D}}S(\rho\|\rho^{*}), \label{eq:qcorr}
\ee
where ${\cal D}$ is a set of all separable states in the Hilbert space, in
which the {\it sub}spaces $A$ and $B$ are also defined. $\rho_A$ and
 $\rho_B$ are the corresponding reduced density matrices, respectively. The
 upper limit of temperatures, at which both quantum and 
 classical correlations exist, is also drawn in Fig.~\ref{phatra} as
 dotted line ($\alpha=0$). As expected, at $T_c$ the quantum correlations
 are zero  meanwhile the classical ones remain finite. That is the case
 also for
 higher temperatures.

\section{Transverse Ising model}
\label{sec:3}

In this section we would like to investigate the quantum correlations
(entanglement) for some spin models in which strong correlations exist and
compare their results with the results of the $SU(2)_c$ colored quark
states given in previous section. The general category of all spin models
goes under the name of the Heisenberg model. For transverse Ising model
with $N$ sites  the Hamiltonian reads
\be 
{\cal H}=-\sum_{i=0}^{N-1}(\lambda\sigma_i^x\sigma_{i+1}^x+\sigma_i^z),
\label{Ham1} 
\ee
where $\sigma^i$, $i=\{x,y,z\}$ are the standard Pauli matrices and
$\lambda$ is the transverse field. It should not be confused with the
eigenvalues given in Eq.~\ref{eq:concur1}. At zero
temperature it is expected that this system goes through a quantum phase
transition QPT (from para- to ferro-magnetic phase transition). 
$\lambda<1$ corresponds to a para-magnetic and $\lambda>1$ 
to a ferromagnetic phase. Unlike the ordinary phase transition, which
occurs at finite temperatures, the fluctuations in QPT are fully quantum
and the transition is due to the changes in the ground state wave
functions. At the critical point long-range correlations are highly
expected. This kind of 
correlations is also expected near the transition point in the
{\it classical} thermal phase transition. However, as the system has zero
temperature and with the assumption that the ground state is
non-degenerate, the system, in whole, is expected to build up a pure
state. It follows that the correlations, which are in principal the
accessible signatures for QPT, are obviously stemming from the long-range
entanglement in the ground state. Through the change in the correlation
functions which will be reflected in fundamental changes in the
entanglement in ground state, we get a distinct signature for probing QPT. \\

\begin{figure}[htb]
\centerline{\psfig{width=10cm,figure=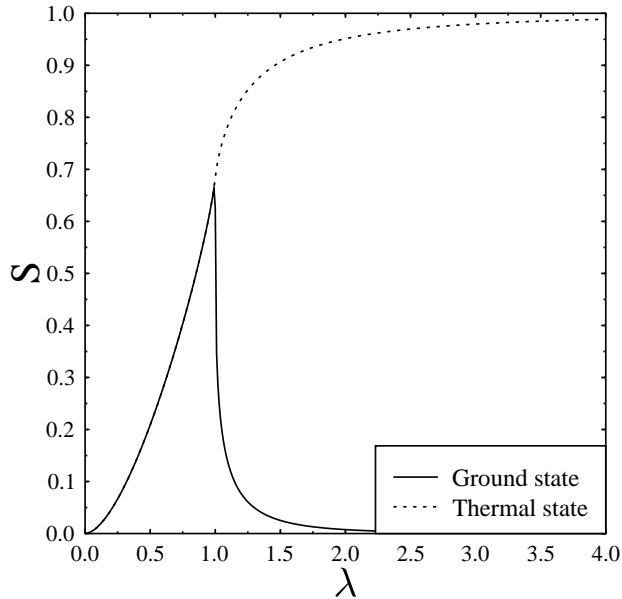}}
\vspace{-2cm}\caption{Zero temperature single-site entanglement $S$ of
  the ground state of transverse Ising model (solid line). The
  dashed line gives the corresponding results for the thermal ground state.
\protect\label{singlesite}}
\end{figure}

\begin{figure}
\centerline{\psfig{width=10cm,figure=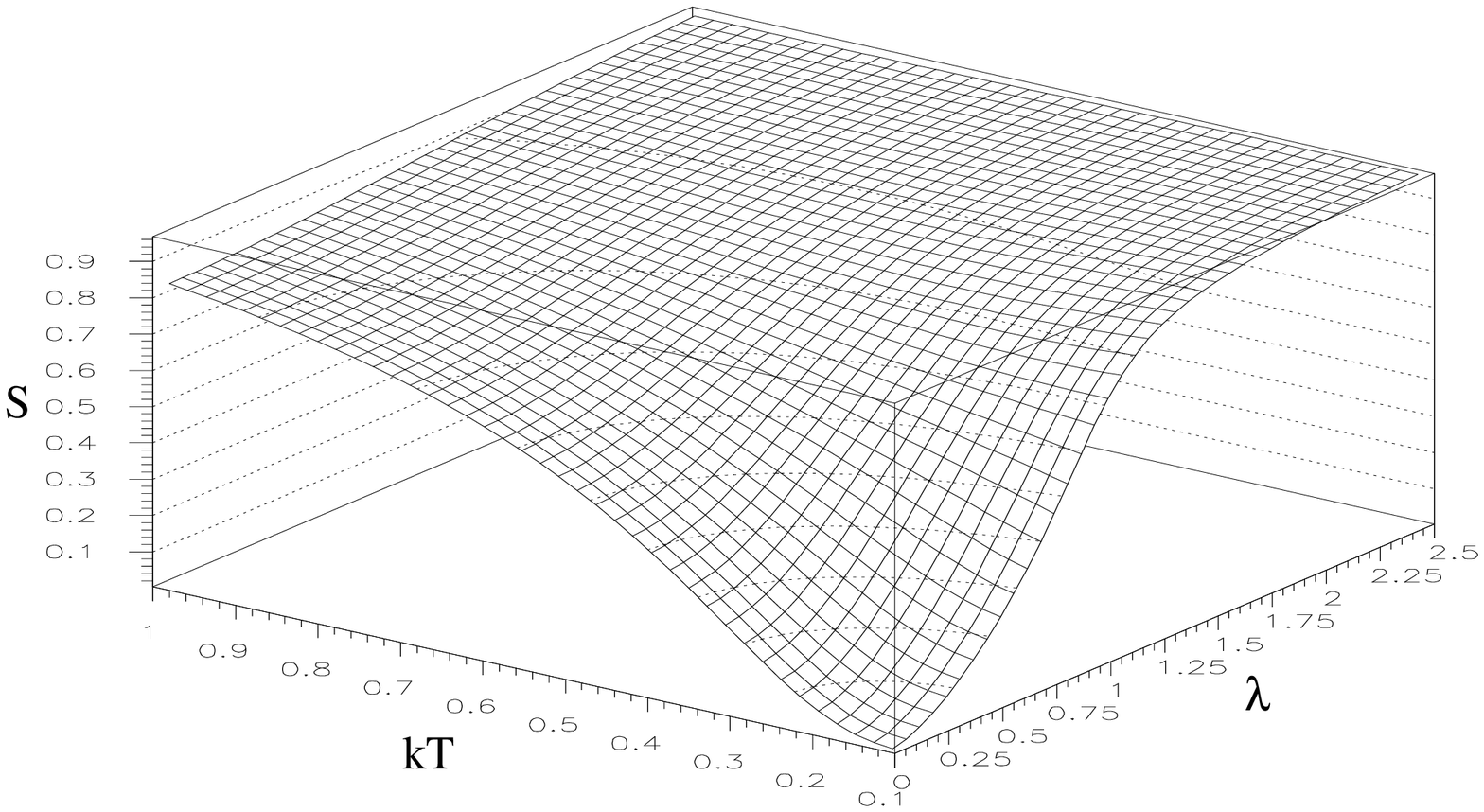}}
\vspace{-.5cm}\caption{Single-site thermal entanglement $S$ as a function of
  $T$ and $\lambda$ in transverse Ising model. 
\protect\label{singlesiteT}}
\centerline{\psfig{width=10cm,figure=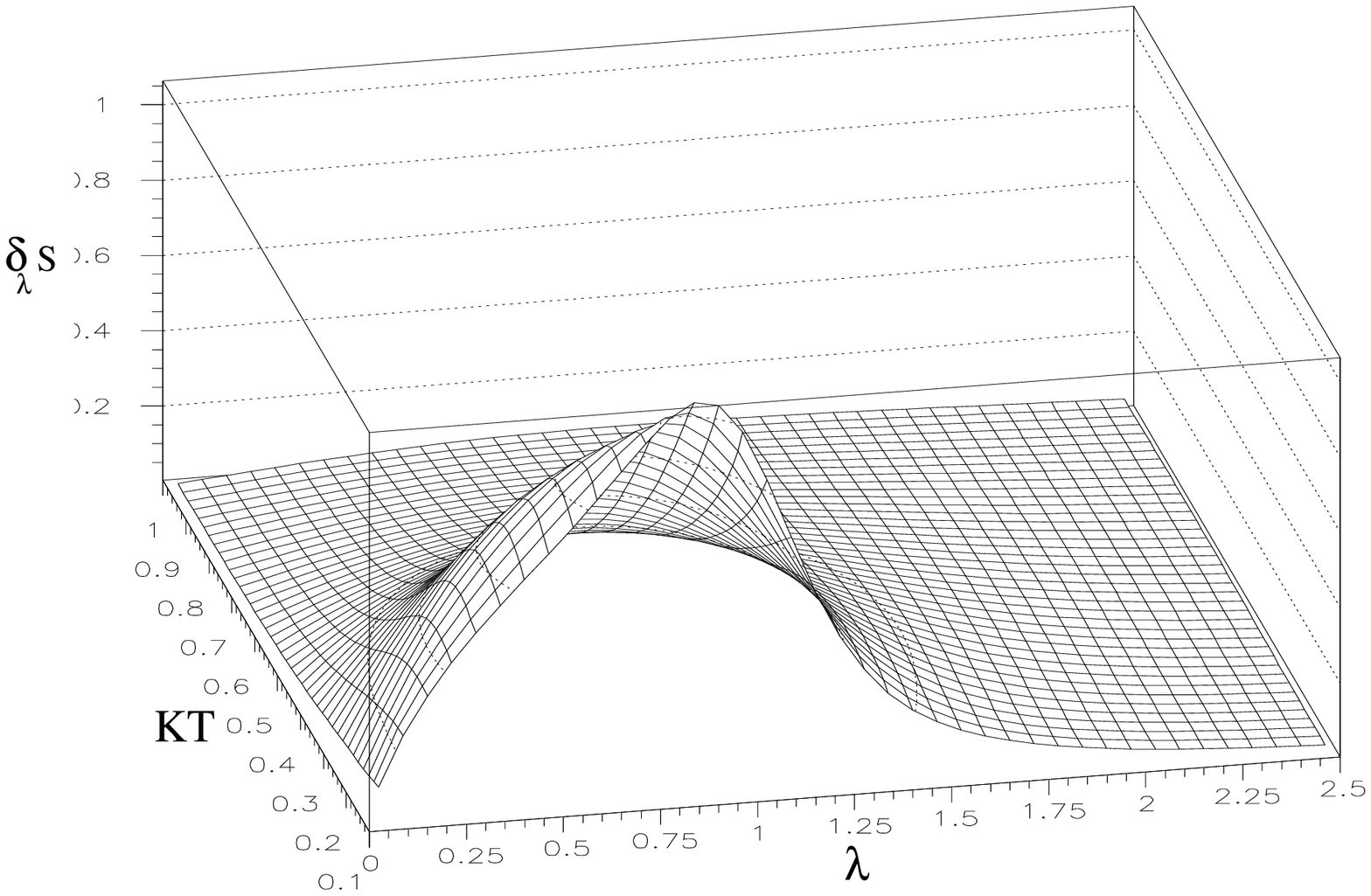}}
\vspace{-.5cm}\caption{First derivative of single-site thermal
  entanglement $\partial_{\lambda} S$ as a function of $T$ and $\lambda$. 
\protect\label{singlesiteT2}}
\end{figure}

\begin{figure}[htb]
\centerline{\psfig{width=10cm,figure=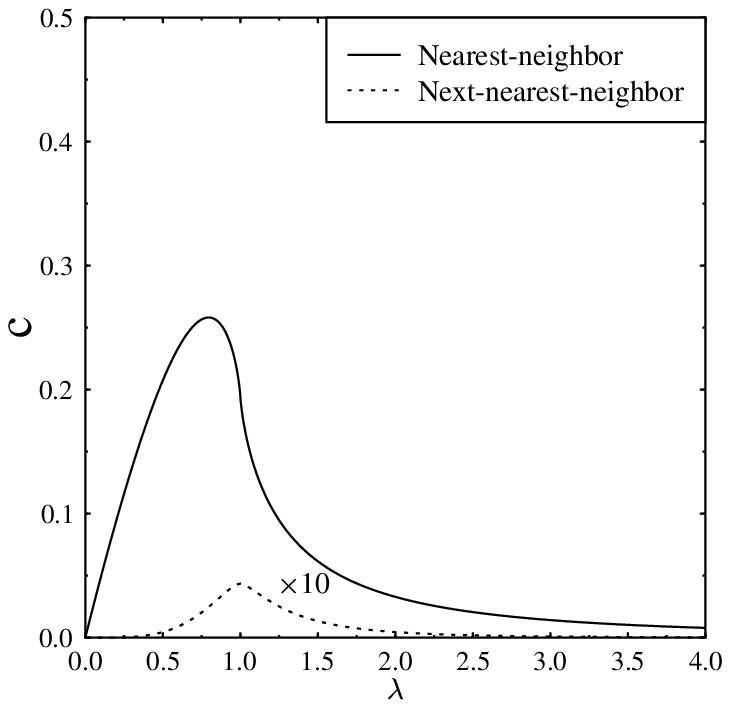}}
\vspace{-2cm}\caption{The entanglement measured by the concurrence of
  nearest- and next-nearest-neighbor two spins at zero temperature depicted
  in dependence on the transverse field $\lambda$. The
  next-nearest-neighbor concurrence is very small. Here it is multiplied by
  factor $10$. 
\protect\label{twosite}}
\end{figure}

Before we study the two-site entanglement we want to know the role
of $\lambda$ in describing the entanglement. To illustrate
this we start in Fig.~\ref{singlesite}  
with the single-spin quantum entropy~\cite{19} which in turn brings  
about a device for measuring the entanglement of one spin with the rest of
the chain. This can be given by evaluating the entropy of the single-site
density matrix $\rho_A$ of the ground state after tracing out all other
spins (see appendix). In transverse Ising model we notice that $S$ varies
from zero at $\lambda=0$ to a maximum value at $\lambda=1$
(Fig.~\ref{singlesite}). For higher values of $\lambda$, $S$ declines,
rapidly, and for $\lambda\rightarrow \infty$, 
$S$ reaches the asymptotic value zero. For the thermal state
($\rho_i=\lim_{T\rightarrow 0}{\rm Tr}_{j\neq i}e^{-\beta {\cal H}}/Z$)  we
can also apply the von~Neumann prescription for calculating the entropy.
We get here the same behavior as in the ground state entropy for
$0\leq\lambda\leq1$. But for $\lambda>1$, we notice 
that $S$ goes on with the increasing until it reaches the
asymptotic value $1$ for $\lambda\rightarrow \infty$. This is because of
the realization of equal mixing of spin thermal states. We can so far conclude
that the single-spin entropy in the thermal ground state is not measuring the
entanglement in the limit $\lambda\rightarrow \infty$. It merely gives
the degree of mixing. For the first time, we study here the single-site thermal
quantum entropy and its derivative with respect to $\lambda$. It should be
understood that in this case the quantum entropy is not a measure for the
entanglement, however one can expect that the change in the quantum entropy
could, in turn, reflect the change in the entanglement. 
In Figs.~\ref{singlesiteT}~and~\ref{singlesiteT2} we show 
the results in $kT-\lambda$ plan. We notice that when $\lambda=1$ both
functions are approximately maximum ($\forall T$). As expected,  the first
derivative of the quantum entropy for high temperatures is always picked at
$\lambda= 1$. This region is not fully compatible with the
thermal concurrence published in~\cite{Osbo02}. Concretely, when we compare
our results (Figs.~\ref{singlesiteT}~and~~\ref{singlesiteT2}) with the
Fig.~6~in~\cite{Osbo02}. In the latter we see that for high temperatures
the concurrence increases for decreasing $\lambda$!

\begin{figure}[tbh]
\centerline{\psfig{width=10cm,figure=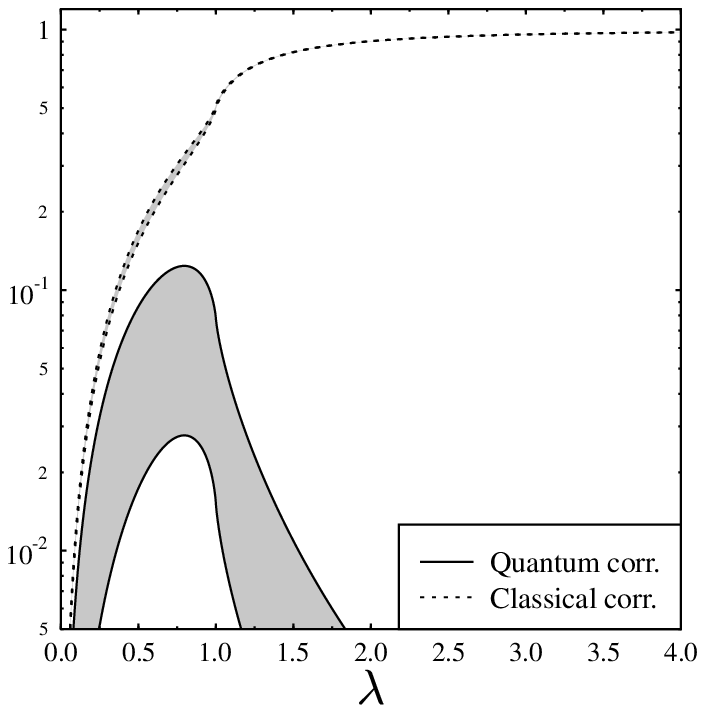}}
\vspace{-1cm}\caption{Bounds on classical and quantum correlations for
  transverse Ising model are drawn versus the strength of the transverse
  magnetic field $\lambda$. 
\protect\label{twositecorr}}
\end{figure}

\begin{figure}[tbh]
\centerline{\psfig{width=10cm,figure=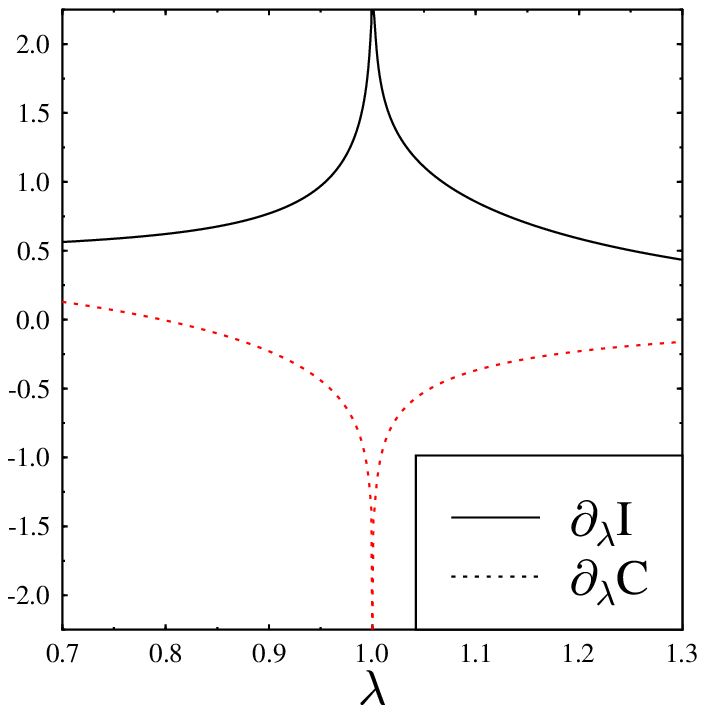}}
\vspace{-2cm}\caption{First derivative of the quantum and classical
  correlations as functions of $\lambda$. For $\lambda<1$ both derivatives
  are finite. But they are vanishing for $\lambda\rightarrow \infty$. At the
  critical point ($\lambda=1$) both functions are divergent, indicating radical
  change in the ground state structure. 
\protect\label{di} 
}
\end{figure}

The two-site entanglement, as its name indicates, measures how two spins
separated by a distance $r$ are entangled. This measure is to be
accomplished by evaluating the concurrence of the entanglement of formation
of two-site density matrix after tracing out all other spins in the chain (see
appendix). The two-site entanglement for nearest-neighbor and
next-nearest-neighbor spin-pairs at zero temperature for the transverse
Ising model are given in Fig.~\ref{twosite}. The length of entanglement
$\xi$  aparently vanishes for all other pairs, since the
concurrence gets negative.  Therefore, the true
non-local quantum part is  a short-range one, meanwhile the
classical part is of long length (diverging for an 
infinite system). We now would like to compare the classical with 
the quantum correlations. We use the same formalism as in the previous
section, however, since it is not possible to directly evaluate the 
relative entropy for the two-site density matrix, instead of this we can bound 
it (upper with lower bounds) to estimate the quantum correlations
and afterwards extract the classical ones. The upper bounds are
calculated from the ${\rm entanglement~of~formation}$  
\be 
E_r^{\rm max}= h\left(\frac{(1+\sqrt{1-{\cal C}^2})}{2}\right) \label{topB}
\ee
where $h(x) = -x\log x -(1-x)\log(1-x)$ is the Shannon entropy. The lower
bounds read~\cite{Vers01}
\be 
E_r^{\rm min} = ({\cal C}-2)\ln\left(1-\frac{{\cal C}}{2}\right) + (1-{\cal
  C})\log(1-{\cal C}) \label{BotB}
\ee
In Fig.~\ref{twositecorr} the bounds on classical and quantum
correlations are depicted in dependence on $\lambda$. We see that the
quantum correlations are smaller than the classical ones. The latter are 
merely produced by the entanglement in the ground state. This suggests the
study of the mutual information as well as the single- and two-site
entanglement~\cite{19,work}, because the mutual
information combines all correlations (classical and quantum) which are
created by the entanglement~\cite{work}. \\

In~\cite{19} the scaling of concurrence has been studied. It has been
found that for infinite system the first derivative of the concurrence
$\partial_{\lambda}{\cal C}$ is a 
diverging function at the critical point. It is important to see
whether the mutual information has such a property.  We show in
Fig.~\ref{di} the first derivative of the mutual information
$\partial_{\lambda}I$, top curve. We obviously get for the first derivative
of concurrence $\partial_{\lambda}{\cal C}$, bottom curve, the same
dependence as we obtained for $\partial_{\lambda}I$. This has been studied
elsewhere~\cite{19}. As expected, at  
the critical point $\lambda=1$ we notice that $\partial_{\lambda}I$ is
singular and divergent. This divergence is a non-ambiguous signal for the
change in the ground state structure. Therefore, we expect that the mutual
information can also be used, in order to probe the entanglement in 
ground state. \\

Thought the color density matrix Eq.~\ref{eq3}, as discussed in the previous
section,  does not undergo QPT, it
is still possible to be compared with the transverse Ising model at zero
temperature. In fact for $\lambda>1$ the mixing due to the change of the
temperature in colored system is compatible with the degeneracy mixing in
the ground state. Clearly, for $\lambda>1$ the relative entropy
(Fig.~\ref{entcla},~\ref{twosite}~and~\ref{twositecorr})  in both systems
are decreasing functions for increasing $T$ and $\lambda$. However, the
classical correlations are different in both systems (see
Fig.~\ref{entcla}~and~\ref{twositecorr}). This is due to the tendency of
the color system to be in a completely random state by loosing all kind of
correlations at high temperatures.

\section{Conclusions and outlook}
\label{sec:4}

In the limit of the proposed {\it thermal} models for colored quarks making up
the hadron physical states we have utilized the concurrence, in order to
study the quantum correlations in $SU(2)_c$ colored quark states. 
The aim of this study is to see whether the quantum and/or the classical
correlations can survive the deconfinement phase transition from hadron to
quark-gluon plasma. Obviously, we have found a region, where an
entanglement transition is highly expected. This gives an upper temperature
limit $T_c$, below which the quantum correlations are significant. At
temperatures higher than $T_c$ the quantum correlations entirely disappear
allowing dominating classical correlations. 

We have also studied the correlations in another system, that  
has the same $SU(2)$ symmetry, namely the transverse Ising model. 
Using the single-site entropy we have identified a certain region in the 
$T-\lambda$ plan, where the quantum effects likely dominate the behavior of
this spin system. The classical correlations can be extracted from the
knowledge of quantum and total correlations. We can conclude that the nature of
quantum correlations is short-ranged, meanwhile that of classical correlations
is long-ranged. We have also shown that the mutual information, which
combines both types of correlations, is a possible candidate to study
the structure of QPT. In a forthcoming work we shall address the mutual
information and its impact in describing the corrlations at finite
temperatures in more details. Also we shall utilize the other models
given in~\cite{Mill03} for the $SU(3)_c$ quark states, which aparently is more
complicated than $SU(2)_c$. It might be possible to compare the quantum
phase trnasition in $SU(3)_c$ colored quark states with the possible
transition in the interior of compact stellar.

\newpage
\section{Appendix}
\label{sec:5}

For the one-dimensional quantum Ising model with $N$ spin-$1/2$ sites the
one- and two-point correlation functions cab be calculated by using the
operator expansion for the total density matrices in terms of the Pauli
matrices.  
\begin{eqnarray}
\rho_i&=&\frac{1}{2}\sum_{k=0}^3\langle\sigma_k^i\rangle\sigma_i^k\, \\
\rho_{k,l}&=&\frac{1}{4}\sum_{i=0,j=0}^3 \langle\;\sigma_k^i
\sigma_l^j\rangle\;\sigma_k^i\otimes\sigma_l^j\,.
\end{eqnarray}
As given in~\cite{Baro00} we can replace the sums by
integrals and afterwards calculate the different correlations in dependence
on the separations between arbitrary two-sites, $r = |k - l|$. Then for the
thermodynamic limit $N\rightarrow\infty$ we find that
\begin{eqnarray} 
\langle\sigma_i^z\rangle &=& -\frac{1}{\pi}\int_0^{\pi}dk(1 + \lambda \cos(k))
       \frac{\tanh(\beta\omega/2)}{\omega} \label{corr1} 
\label{corr2}
\end{eqnarray} 
where $\beta=1/T$. For $T=0$ we have 
\begin{eqnarray} 
\langle\sigma_i^x\rangle &=& \left\{\begin{array}{ll}
       0, & \lambda\leq1\\
       (1-\lambda^{-2})^{1/8},&\lambda>1
       \end{array}\right.
\end{eqnarray} 
Using the definition:
\begin{eqnarray}
C_r&=&\frac{1}{\pi}\int_0^{\pi}dk \cos(kr)(1 + \lambda \cos(k))
 \frac{\tanh(\beta\omega/2)}{\omega} \nonumber \\
&-& \frac{\lambda}{\pi}\int_0^{\pi}dk
 \sin(kr)\sin(k)  \frac{\tanh(\beta\omega/2)}{\omega}\,,
\end{eqnarray}
then the two-site correlation functions read~\cite{Baro00,19}
\begin{eqnarray}
\langle\sigma_0^x\sigma_{r}^x\rangle &=& \left|\begin{array}{cccc}
    C_{-1}&C_{-2}&\dots&C_{-r}\\
    C_{0}&C_{-1}&\dots&C_{-r+1}\\
    \vdots&\vdots&\ddots&\vdots\\
    C_{r-2}&C_{r-3}&\dots&C_{-1}
  \end{array}\right|\,, \label{eq:sigmax}\\
\langle\sigma_0^y\sigma_{r}^y\rangle &=&\left|\begin{array}{cccc}
    C_{1}&C_{0}&\dots&C_{-r+2}\\
    C_{2}&C_{1}&\dots&C_{-r+3}\\
    \vdots&\vdots&\ddots&\vdots\\
    C_{r}&C_{r-1}&\dots&C_{1}
  \end{array}\right| \,, \label{eq:sigmay}\\
\langle\sigma_0^z\sigma_{r}^z\rangle &=&
   \langle\sigma_0^z\rangle^2-C_rC_{-r} \label{eq:sigmaz}
\end{eqnarray}
It should be noticed that in Eq.~\ref{eq:sigmaz} there is no factor $4$ in
the front of the first term as the case in~\cite{19}. As in~\cite{Baro00}
we find that the factor $4$ should be multiplied by the magnetization $m_z$
and not by the polarization $\langle\sigma_z\rangle$. 
Since $m_z=\langle\sigma_z\rangle/2$, we get factor $1$.

\end{document}